\documentclass[aps,prd,amsmath,showpacs,showkeys,preprint]{revtex4}
\usepackage{amsmath}
\usepackage[cp1251,koi8-r]{inputenc}
\usepackage[english]{babel}
\usepackage{graphics}
\usepackage[dvips]{color}
\usepackage{epsfig}
\newcommand{\be}{\begin{equation}}
\newcommand{\ee}{\end{equation}}
\newcommand{\bn}{\begin{eqnarray}}
\newcommand{\en}{\end{eqnarray}}
\newcommand{\bd}{\begin{displaymath}}
\newcommand{\ed}{\end{displaymath}}
\newcommand{\bnn}{\begin{eqnarray*}}
\newcommand{\enn}{\end{eqnarray*}}
\newcommand{\bs}{\begin{subequations}}
\newcommand{\es}{\end{subequations}}

\def\Journal#1#2#3#4#5#6{#1, \ #2, \ #3 \  #4 \ (#5) \ #6.}

\begin{document}
\inputencoding{cp1251}
\title{Regular and stochastic behavior of Parkinsonian pathological tremor signals}
\author{\firstname{R.~M.}~\surname{Yulmetyev}$^{1,2}$}
\email{rmy@theory.kazan-spu.ru}
\author{\firstname{S.~A.}~\surname{Demin}$^{1,2}$}
\email{sergey@theory.kazan-spu.ru}
\author{\firstname{O.~Yu.}~\surname{Panischev}$^{1,2}$}
\affiliation{$^1$Department of Physics, Kazan State University,
420008 Kazan, Kremlevskaya Street, 18 Russia \\ $^2$Department of
Physics, Kazan State Pedagogical University, 420021 Kazan,
Mezhlauk Street, 1 Russia}
\author{Peter H\"anggi}\affiliation{Department of
Physics, University of Augsburg, Universit\"atsstrasse 1, D-86135
Augsburg, Germany}
\author{\firstname{S.~F.}~\surname{Timashev}}
\affiliation{Karpov Institute of Physical Chemistry, 105064
Moskow, Vorontsovo pole Street, 10 Russia}
\author{\firstname{G.~V.}~\surname{Vstovsky}}
\affiliation{Semenov Institute of Chemical Physics, 117977 Moscow,
Kosygina Street, 4 Russia}

\begin{abstract}
Regular and stochastic behavior in the time series of Parkinsonian
pathological tremor velocity is studied on the basis of the
statistical theory of discrete non-Markov stochastic processes and
flicker-noise spectroscopy. We have developed a new method of
analyzing and diagnosing Parkinson's disease (PD) by taking into
consideration discreteness, fluctuations, long- and short-range
correlations, regular and stochastic behavior, Markov and
non-Markov effects and dynamic alternation of relaxation modes in
the initial time signals. The spectrum of the statistical
non-Markovity parameter reflects Markovity and non-Markovity in
the initial time series of tremor. The relaxation and kinetic
parameters used in the method allow us to estimate the relaxation
scales of diverse scenarios of the time signals produced by the
patient in various dynamic states. The local time behavior of the
initial time correlation function and the first point of the
non-Markovity parameter give detailed information about the
variation of pathological tremor in the local regions of the time
series. The obtained results can be used to find the most
effective method of reducing or suppressing pathological tremor in
each individual case of a PD patient. Generally, the method allows
one to assess the efficacy of the medical treatment for a group of
PD patients.
\end{abstract}

 \maketitle

\section{Introduction. Parkinson's disease}

Recently, much effort has been made in searching new alternative
methods of diagnosing, treating and preventing severe diseases of
central nervous and locomotor systems. Among them, Parkinson's
disease (PD) is one of the most serious illnesses. PD, was called
so by the French neurologist Pierre Marie Charcot in the 19th
century to honor Dr. James Parkinson, who first described the
disease in 1817. Dr. Parkinson presented the account of the
observation results made about six patients in his book {\it An
essay on the shaking palsy}.

The present paper deals with two physical methods used in
combination to analyze, diagnose and treat PD. The possibilities
of the methods are assessed and compared. The comparison allowed
us to add extra information about the behavior of PD pathological
tremor physical parameters.

The first method is based on the notions and concepts of the
statistical theory of discrete non-Markov stochastic processes
\cite{Yulm1,Yulm2}. The method is connected with the studies of
statistical non-Markov effects, long- and short-range statistical
memory effects, regularity and stochastic behavior effects, and
dynamic alternation of relaxation modes in the patient in various
dynamic states. The study of non-Markov effects in complex systems
in biology \cite{Goychuk1,Goychuk2,Goychuk3,Gamm}, physics
\cite{Gamm,Mokshin1,Mokshin2,Mokshin3}, seismology
\cite{Yulm3,Yulm4}, and medicine \cite{Yulm2,Yulm5,Yulm6,Yulm7} is
of special interest for correlation analysis. The scale of time
fluctuations, long-range effects, discreteness of various
processes and states, and the effects of dynamic alternation in
the initial time series are important role in this respect. The
discreteness of experimental data, statistical effects of long-
and short-range memory and the constructive role of fluctuations
and correlations can be used to obtain information about the
properties and parameters of the system under study. Within this
method, a set of quantitative and qualitative parameters allows
one to determine typical distinctions between the natural and
aftertreatment states of a patient, describes a detailed variation
in patient's pathological tremor, and helps to choose the most
effective treatment of separate patients and statistical groups.

The second method, flicker-noise spectroscopy (FNS), is a general
phenomenological approach to the analysis of the behavior of
complex nonlinear systems. It is designed to extract the
information contained in chaotic signals of various natures: time
series, spatial series, and complex power spectra produced by the
systems. Within the FNS method, the series of various
irregularities (spikes, jumps, discontinuities of derivatives of
various orders) of the dynamic variables of a system at all levels
of its temporal and spatial hierarchy are analyzed to extract the
information that can help to predict the behavior of the system.
The use of irregularities in the dynamic variables as the
information basis or "colors" of the FNS methodology enables us
not only to classify all the information contained in the chaotic
series in the most general phenomenological form, but also to
extract distinctively any desired portion of this information. The
FNS method can be used to solve three types of problems:  to
determine the parameters, or patterns, characterizing the behavior
or structural features of open complex (physical, chemical,
natural) systems; to determine the precursors of the sharpest
changes in the state of open dissipative systems of various
natures using the prior information about the behavior of the
systems; to determine the redistribution behavior of excitations
in the distributed systems by analyzing dynamic correlations in
chaotic signals measured simultaneously at different points.

PD is a progressing chronic brain disease manifesting itself in
movement disorders. PD is caused by complex biochemical processes
accompanied by the lack of the chemical substance dopamine.
Dopamine acts as a transmitter of signals from one nervous cell to
another. The lack of the neurotransmitter dopamine causes changes
in the brain parts that control human motor functions.

The origin of PD is not clear. It is most frequently believed that
this disease is caused by a combination of three factors:
biological aging, heredity and exposure to some toxins. The major
PD symptoms are hypokinesia, stiffness and tremor. New ideas and
principles are required to solve the problems emerging in
classical treatment methods. The diagnostics of the disease in the
early stages is one of these problems. Only the joint efforts of
experts from different science areas can bring a solution to this
kind of problems. For example, the modern methods of biophysics,
biochemistry and neurophysiology enabled us to make considerable
progress in understanding PD reasons and developing the methods of
its treatment.

Today, there are three methods of PD treatment: medicamentous
therapy, neurologic surgery and  electromagnetic stimulation of
certain brain areas. To obtain a more reliable picture of PD
evolution, it is necessary to use physical methods. In this case,
statistical methods of analysis are of special importance. They
are used to analyze experimental time series of various parameters
of tremor in patient's limbs. As a rule, the experimental data are
obtained by conventional biomechanical methods
\cite{Smol,Win,Holt} such as recording of electric signals in leg
muscles when the patient is walking \cite{Ding} or by laser
recording of physiological and pathological tremor in human hands
\cite{Timm1,Timm2,Timm3,Timm4,Sapir}.

A nonlinear dynamical model is used to analyze a variety of human
gaits \cite{West}. The stride-interval time series in normal human
gait is characterized by slightly multifractal fluctuations. The
fractal nature of the fluctuations becomes more pronounced under
both the increase and decrease in the average gait \cite{West}.

Nonlinear time series analysis is applied in studying normal and
pathological human walking \cite{Ding}. The problems caused by age
changes in a human gait \cite{Haus1, Haus2, Haus3, Haus4}, various
movement disorders and locomotor system diseases \cite{Haus5,
Haus6} are studied by J. Hausdorff. A nonlinear signal, multimodal
(independent) oscillations, and the periodic pattern of time
records in hand tremor and muscle activity in a PD patient are
studied in \cite{Sapir}. PD patients exhibit tremor, involuntary
movements of the limbs. Typically the frequency spectrum of tremor
has broad peaks at "harmonic" frequencies, much like that is seen
in other physical processes. In general, this type of harmonic
structure in the frequency domain may be due to two possible
mechanisms: a nonlinear oscillation or a superposition of
(multiple) independent modes of oscillation \cite{Sapir}. Various
dynamical states of PD patients are also studied by means of the
time series of pathological tremor in fingers
\cite{Beuter1,Beuter2,Beuter3,Vaill1,Vaill2}.

At present special attention is attached to the problems of
distinguishing and analyzing of the stochastic and regular
components of experimental time series of biological systems.
Thereto various methods of nonlinear physics and simulation by
nonlinear oscillators \cite{Babl1,Babl2,Gold1,Gold2,Gold3}, the
methods of fractal analysis of time series \cite{Lieb1,Lieb2,
Lieb3,Lieb4,Lieb5,Lieb6}, the methods of detrended fluctuation
analysis \cite{Peng1,Peng2} et al. are used.

In Ref. \cite{Babl1} Babloyantz \emph{et al.} presented a simple
graphical method that unveils subtle correlations between short
sequences of a chaotic time series. Similar events, even from
noisy and nonstationary data, are clustered together and appear as
well-defined patterns on a two-dimensional diagram and can be
quantified. The general method is applied to the electrocardiogram
of a patient with a malfunctioning pacemaker, the residence times
of trajectories in the Lorenz attractor as well as the logistic
map. In this paper the authors introduced a simple graphical
method that projects the trajectories of chaotic attractors into
various planes in a way mat unveils very subtle correlations
between consecutive sequences of events. The merit of the method
resides in the fact that these sequences may extend over more than
ten events. The sequences with similar relationships but not the
same absolute values may appear as well-defined structures in a
two-dimensional diagram. The advantage of this mapping is that,
although the sequences may take into account more than three
consecutive events, the two-dimensional projections are extremely
helpful visual aides for elucidation of some aspects of chaotic
dynamics.

Goldbeter \emph{et al.} study various biological rhythms which
regulate the vital activity of living systems and determine the
control mechanisms of these rhythms. In Ref. \cite{Gold1} the
periodic oscillations at all levels of biological organization,
with periods ranging from a fraction of a second to years were
observed. The authors of Ref. \cite{Gold2} examined the mechanisms
of transitions from simple to complex oscillatory phenomena in
metabolic and genetic networks. The mechanisms underlying such
transitions are examined in models for a variety of rhythmic
processes in several live systems. Ref. \cite{Gold3} is devoted to
the study of circadian oscillations stochastic dynamics and the
influences the gene expression exerts on it. Authors show the way
robust circadian oscillations are produced from a "bar-code"
pattern of gene expression.

Liebovitch \emph {et al.} reveal the local features of dynamics of
various biological systems by the methods of fractal analysis and
various methods of nonlinear physics. In Ref. \cite{Lieb1} the
question was examined: the way the dynamics of neural networks of
the Hopfield type depends on the updating scheme, temperature
dependence, degree of locality of connections between elements and
the number of memories. Further the results were applied to
interpret some features of protein dynamics. In Refs. \cite{Lieb2,
Lieb3} authors studied the self-organizing and the synchronization
of the trajectory of a coupled systems dynamics. In particular, in
Ref. \cite{Lieb2} the authors introduced a scheme of controlling
the dynamics of deterministic systems by coupling it to the
dynamics of other similar systems. The controlled systems
synchronized their dynamics with the control signal in periodic as
well as chaotic regimes. In Ref. \cite{Lieb4} the processes of
transition from persistent to antipersistent correlation was
studied by means of fractal analysis methods of a time series:
fractional Brownian motion, rescaled range analysis, variance
analysis, zero-crossing analysis. The authors discussed several
simple random walk models which produce such transitions (bounded
correlated random walk, fractional Brownian motion with a long
relaxation time), and therefore are candidates for the mechanisms
that may be present in some biological systems. The authors have
studied the way the pattern, seen in the experimental data of
biological systems, persistent at short time intervals and
antipersistent at long time intervals, could arise from dynamical
systems. The comparison of Hurst coefficient, which was calculated
on the basis of different fractal analysis methods, was carried
out. The article \cite{Lieb5} is devoted to the study of the
cardiac rhythm abnormalities by means of estimation of the
probability density function and Hurst rescaled range analysis. In
paper \cite{Lieb6} the authors examined the effects of fractal
ion-channel activity in modifications of two classical neuronal
models: Fitzhugh-Nagumo and Hodgkin-Huxley. The authors came to
the conclusion that fractal ion channel gating activity was
sufficient to account for the fractal-rate firing behavior.

In this paper a qualitatively new methodology of extracting the
information from the time series on the united basis of the theory
of discrete non-Markov stochastic processes and the flicker-noise
spectroscopy for the case of Parkinson's disease is submitted.
This methodology presents a simple graphic and relatively
inexpensive method of analysis of various physiological and
pathological patient's states. In particular, it allows to make a
quantitative estimation of the quality of treatment and to define
the most effective method of treatment in each individual case and
for a group of patients. We determine the structure of the initial
time signal, and also the information about the non-stationary
effects or about the dynamic alternation by a set of quantitative
physical characteristics. We give special attention to the
analysis of correlations and fluctuations which determine the time
evolution in live systems.

In particular, the signals of PD pathological tremor are
physically interpreted to answer the following questions:

(i) How can the study of the stochastic behavior and regularity in
tremor signals help in evaluating the state of a PD patient?

(ii) How do certain physical parameters related to Markov and
non-Markov features, statistical memory effects and dynamic
alternation of relaxation modes in the initial time signal change?

(iii) How do the low- and high-frequency components of the initial
time signal respond to the changes of pathological tremor in the
patient under treatment?

\section{Basic concepts and definition of
statistical theory of discrete non-Markov random processes}

The theory of discrete non-Markov stochastic processes
\cite{Yulm1, Yulm2} is based on the finite-difference
representation of the kinetic Zwanzig-Mori's equations
\cite{Zwanzig1,Zwanzig2,Mori1,Mori2} for condensed matters, which
are well known in the statistical physics of nonequilibrium
processes. The theory is also widely used in analyzing complex
biological and social systems. Dynamic, kinetic and relaxation
parameters provided by this theory contain detailed information on
a wide range of parameters and properties of complex systems.

Let the behavior of PD pathological tremor velocity be described
by a discrete time series $x_j$ of variable $X$: \be
X=\{x(T),x(T+\tau),x(T+2\tau),...,x(T+\tau N-\tau)\}. \ee Here $T$
is the time at which the recording of the pathological tremor is
started, $(N-1)\tau$ is the total time of signal recording, and
$\tau$ is the discretization time. In the system under study the
discretization time is $\tau=10^{-2}$ sec. To describe the dynamic
parameters of pathological tremor (correlation dynamics), it is
convenient to use a normalized time correlation function (TCF):
\be \nonumber a(t)=\frac{1}{(N-m)\sigma^2}\sum_{j=0}^{N-1-m}\delta
x_j\delta x_{j+m}=\ee \be =
\frac{1}{(N-m)\sigma^2}\sum_{j=0}^{N-1-m}\delta x(T+j\tau)\delta
x(T+(j+m)\tau),  \ee \be \nonumber t=m\tau,~~ 1 \leq m \leq N-1.
\ee TCF depending on current $t=m\tau$ can be conveniently used to
analyze dynamic properties of complex systems. TCF usage means
that the developed method is true of complex systems, when
correlation function exists. The mean value $\langle X \rangle$,
fluctuations $\delta x_j$, absolute ($\sigma^2$) and relative
($\delta^2$) dispersion for a set of random variables (Eq. 1) can
be easily found by: \be \nonumber \langle X \rangle =
\frac{1}{N}\sum_{j=0}^{N-1} x(T+j\tau), \ee \be \nonumber
x_j=x(T+j\tau), \delta x_j = x_j-\langle X \rangle, \ee \bn
\nonumber \sigma^2 = \frac{1}{N}\sum_{j=0}^{N-1}\delta x_j,
\delta^2 = \frac{\sigma^2}{\langle X \rangle^2}. \en
 The function $a(t)$ satisfies the normalization
and relaxation conditions of correlations: $\lim_{t \to 0}a(t)=1,
~\lim_{t \to \infty}a(t)=0$.

By using the Zwanzig-Mori's technique of projection operators
\cite{Zwanzig1,Zwanzig2,Mori1,Mori2} it is possible to receive an
interconnected chain of finite-difference equations of a
non-Markovian type \cite{Yulm1,Yulm2} for the initial TCF $a(t)$
and the normalized memory functions in the following way: \bn
\frac{\triangle a(t)}{\triangle
t}=\lambda_1a(t)-\tau\Lambda_1\sum_{j=0}^{m-1}M_1(j\tau)a(t-j\tau),
\\  \ldots \nonumber \en Here $\lambda_1$ is the eigenvalue and $\Lambda_1$ is the
relaxation parameter of Liouville's quasioperator $\hat{L}$.
Function $M_1(j\tau)$ is a normalized memory function of the first
order: \be \nonumber \lambda_1=i\frac{\langle
\textbf{A}_k^0(0)\hat{L}\textbf{A}_k^0(0)\rangle}{\langle|\textbf{A}_k^0(0)|^2\rangle},
~~\Lambda_1=\frac{\langle
\textbf{A}_k^0(0)\hat{L}_{12}\hat{L}_{21}\textbf{A}_k^0(0)\rangle}{\langle|\textbf{A}_k^0(0)|^2\rangle},
\ee \be M_1(j\tau)=\frac{\langle
\textbf{A}_k^0(0)\hat{L}_{12}(1+i\tau\hat{L}_{22})^j
\textbf{A}_k^0(0) \rangle}{\langle
\textbf{A}_k^0(0)\hat{L}_{12}\hat{L}_{21}\textbf{A}_k^0(0)\rangle},~~
M_1(0)=1. \ee

Gram-Schmidt orthogonalization procedure $\langle \textbf{W}_n,
\textbf{W}_m\rangle=\delta_{n,m}\langle |\textbf{W}_n|^2\rangle$,
where $\delta_{n,m}$ is Kronecker's symbol, can be used to rewrite
the above equations in a compact form: \be \nonumber
\textbf{W}_0=\textbf{A}_k^0(0),~~\textbf{W}_1=(i\hat{L}-\lambda_1)\textbf{W}_0,~~\textbf{W}_2
=(i\hat{L}-\lambda_2)\textbf{W}_1-\Lambda_1\textbf{W}_0,...,\ee
\be \nonumber
\textbf{W}_n=(i\hat{L}-\lambda_n)\textbf{W}_{n-1}-\Lambda_{n-1}\textbf{W}_{n-2}-...~.\ee
Then the eigenvalue $\lambda_{1}$ of Liouville's quasioperator and
the relaxation parameter $\Lambda_{1}$ in Eq. (3) take the form
of: \be \nonumber \lambda_{1}=i\frac{\langle
\textbf{W}_{0}\hat{L}\textbf{W}_{0} \rangle}{\langle
|\textbf{W}_{0}|^2\rangle}, ~~ \Lambda_{1}=i\frac{\langle
\textbf{W}_{0}\hat{L}\textbf{W}_{1} \rangle}{\langle
|\textbf{W}_{0}|^2\rangle}. \ee The normalized memory function of
the first order in Eq. (3) is rewritten as: \be \nonumber
M_1(t)=\frac{\langle
\textbf{W}_1(1+i\tau\hat{L}_{22})^m\textbf{W}_1\rangle}{\langle
|\textbf{W}_1(0)|^2 \rangle}. \ee

The finite-difference kinetic equation (3) represents the
generalization of Zwanzig-Mori's kinetic theory
\cite{Zwanzig1,Zwanzig2,Mori1,Mori2}, which is well known in
statistical physics, for complex discrete non-Hamiltonian
statistical systems. Within our method of the analysis of dynamics
of the statistical time series we do not use the equation (3) as
an object for the subsequent theoretical analysis. In this
connection we do not use the equations such as Zwanzig-Mori's for
memory functions of the 2-nd and higher orders. We use the
algorithm, which was above described, to calculate the time
dynamics $a(t)$, $M_1(t)$ and parameters $\lambda_1$, $\Lambda_1$.
The dependences $a(t)$ and $M_1(t)$ are calculated on the basis of
the experimental data, independently of each other. At the same
time we control the conformity of the calculated dependences
$a(t)$, $M_1 (t)$ and parameters $\lambda_1$, $\Lambda_1$ to the
equation (3) (the precision of the conformity is $\sim 2 - 5 \%$
for the cases described here). We use the dependences $a (t)$ and
$M_1 (t)$ to analyze the amplitude of Parkinsonian tremor
velocity. We also use these dependences to calculate the
non-Markovity parameter \cite{Yulm1, Yulm2} which characterizes
the degree of correlativity of the signal. The studies, which have
been carried out earlier \cite{Yulm2, Yulm5, Yulm6}, show that
this parameter contains detailed information about the
physiological state of a system.

In this paper we shall use the spectral dependence
$\varepsilon_1(\nu)$ of the first point of the non-Markovity
parameter \cite{Yulm1, Yulm2}: \be \varepsilon_1(\nu) = \left\{
\frac {\mu _ {0} (\nu)}{\mu_1 (\nu)} \right\}^{\frac {1} {2}}, \ee
which is determined by means of Fourier transformations
$\mu_0(\nu)$, $\mu_1(\nu)$ of functions $a(t)$ and $M_1(t)$
respectively: \be \nonumber
\mu_0(\nu)=\left|\sum_{j=0}^{N-1}a(t_j)\cos(2\pi\nu
t_j)\right|^2,~~\mu_1(\nu)=\left|\sum_{j=0}^{N-1}M_1(t_j)\cos(2\pi\nu
t_j)\right|^2. \ee Further we shall show, that the application of
the frequency-dependence $\varepsilon_1(\nu)$ and the values of
this parameter on zero frequency: \be
\varepsilon_1(\nu=0)=\varepsilon_1(0)=\left\{ \frac {\mu _ {0}
(0)}{\mu_1 (0)} \right\}^{\frac {1} {2}}, \ee allows to introduce
the quantitative estimations for various dynamic states of a
patient with PD. In particular, we shall show, that the values of
parameter $\varepsilon_1(0)\sim 10^1$ for the analyzed system are
characteristic of stable physiological states (for the patient
under treatment). The appearance of pathology in a system leads to
a sharp decrease in this parameter, approximately by one order.
Thus, we can compare quantitatively  various dynamic states of the
studied system considering the changes of the non-Markovity
parameter.

\section{Flicker-noise spectroscopy for analysis of time series of dynamic variables}
New information about chaotic time signals can be obtained by the
method of the flicker-noise spectroscopy (FNS)
\cite{Timsh01,Timsh02,Timsh03,Timsh04,Timsh05}. Its advantage
consists in extracting information from the series of distinct
irregularities (spikes, jumps, discontinuities of derivatives of
various orders) by analyzing the behavior of time, spatial and
power dynamic variables at each existential level of the
hierarchical organization of the system. Thus, the most valuable
information is obtained by analyzing the power spectra and the
difference moments ("structural functions") of various orders. It
is necessary to point out, that the difference moments are formed
exclusively by irregularities of a jump type. On the other hand
the power spectra are formed by the contributions of two types of
irregularities: peaks and jumps.

The FNS method was applied to analyze the dynamics of various
physical and chemical processes
\cite{Timsh06,Timsh07,Timsh08,Timsh09,Timsh10,Timsh11,Timsh12,Timsh13}.
Among them are fluctuations of electric voltage in electrochemical
systems (considered in the process of formation of porous silicon
under conditions of anodic polarization, formation of molecular
hydrogen on platinum under cathodic polarization, initiation of
hydrodynamical instability in the field of an over-limiting
current in electro-membrane systems), fluctuating dynamics of the
solar activity, fluctuation of a velocity component in turbulent
streams. Unique abilities of the FNS method to locate the
multipoint correlation interrelations were shown in works
\cite{Timsh14, Timsh15,Timsh16}. That was done on the basis of the
analysis of simultaneously measured signals at spaced points of
the distributed systems.

The basic relations of FNS are given below. We analyze the chaotic
series of dynamic variable $x(t)$ over the time interval
$T_{tot}=N\tau$, where $\tau$ is a sampling time.

1. We proceed from the notion of hierarchy of spatial-temporal
levels of the organization of open dynamic dissipative systems.

2.  The most valuable information in the chaotic series is stored
in irregularities of various types, such as peaks, jumps, breaks
of derivatives of different orders. The parameters which
characterize the properties of irregularities, can be obtained by
the analysis of power spectra $S(f)$ ($f$ is a frequency): \be
S(f)=\left[ \frac{1}{N}\sum_{k=1}^{N} \delta x_k \cos(2\pi f k
\tau) \right]^2 +\left[ \frac{1}{N}\sum_{k=1}^{N} \delta x_k
\sin(2\pi f k \tau) \right]^2,\ee where $x_k=x(T+t_k)$ and
$t_k=k\tau$, \be \nonumber \delta x_k=x_k-\frac{1}{N} \sum_{k=1}^N
x_k, \ee and also by the analysis of transitive difference moments
("transitive structural functions") $\Phi^{(2)}(t_n)$ of the
second order: \be \Phi^{(2)}(t_n)=\frac{1}{N-n}\sum_{k=1}^{N-n}
\left[ x(t_k)-x(t_{k+n})\right]^2, \ee where $t_n$ is a delay
parameter. Further, when considering the dependence
$\Phi^{(2)}(t_n)$ we will not specify the bottom index.

It should be noted that the proposed averaging procedure differs
from Gibbs's procedure when averaging is carried out by using the
probability density. Actually, we do not consider the statistics
of ensembles as it is done in the statistical Gibbs's
thermodynamics which is based on ergodic hypothesis. We also
generalize Einstein's approach to the analysis of fluctuation
dynamics \cite{Timsh17,Timsh18}.

3. Parameters or "passport data", obtained by the analysis of
dependences $S(f)$ and $\Phi^{(2)}(t_n)$, are correlation times
and dimensionless parameters. These dimensionless parameters
describe the loss of "memory" (correlation relations) in
irregularities of a "spikes" and "jumps" type.

4. For stationary processes in open dissipative systems the moment
$\Phi^{(2)}(t_n)$ depends only on the difference of arguments
$n\tau$. The self-similar structure is realized in this case. It
means that the dependences $S(f)$ or $\Phi^{(2)}(t_n)$ are
identical for each level of the system hierarchichy.

It should be noted, that  the reverse transformations in the FNS
methodology are not used in the way it takes place in Fourier- or
in wavelet-analysis. Therefore, no constraints are imposed on the
character of the dependence $x(t)$ except for the existence of
average values.

\subsection{Basic equations for stationary processes}

Let's obtain approximations for $S (f)$ and $\Phi^{(2)}(t_n)$,
which are determined by irregularities in the behavior of dynamic
variables. At the first stage the generalized $\delta$-functions
are used to approximate the spikes of the dynamic variables, and
Heavyside functions are used to approximate the jumps. At the same
time the "low-frequency" limit $f \ll 1/2 \pi T_i$ is considered,
when the characteristic time intervals $T_i$ between the nearest
irregularities are much less than all the characteristic times of
the considered system. In case of stationary processes the
obtained expressions are the same for each of the hierarchical
levels. Then simple approximation dependence (see
\cite{Timsh01,Timsh02,Timsh03,Timsh04,Timsh05} for more
information) can be obtained for $S(f)$ and $\Phi^{(2)}(t_n)$.

The approximation for the structural function of the second order
reads: \be \Phi^{(2)} (t)=2\cdot\sigma^2 \cdot \left[
1-\Gamma^{-1} (H_1)\cdot \Gamma (H_1,t/T_1)\right]^2.\ee Here
$\Gamma (s)$ and $\Gamma(s, x)$ are respectively the
gamma-function and incomplete gamma-function $(x \geq 0$ and $s
> 0)$; $\sigma$ is dispersion of the measured dynamic variable
with dimension $[x]$. Value $H_1$ is a Hurst's parameter. It
characterizes the rate of loss of "memory" during the time
intervals shorter than the correlation time $T_1$. As follows from
Eq.(9), $\Phi^{(2)}(t) \rightarrow \sigma^2$ for $t \gg T_1,$
parameter $T_1$ actually characterizes the time interval during
which "forgetting" of the previous value of the dynamic variable
occurs. Such "forgetting" is the consequence of the "jumps" of the
dynamic variable at each level of the spatio-temporal hierarchy.
True, the structural function is zero, $\Phi^{(2)}(t) = 0$, for
the sequences of the irregularities-spikes, which are represented
by a sequence of $\delta$-functions (see chapter 4.3 (Fig. 51) of
Schuster's monograph \cite{Timsh18}). At the same time, a
continuous power spectrum $S(f)$ can be calculated for the
correlated sequence of $\delta$-functions within a low-frequency
limit at $f \ll 1/2\pi T_i$, as was shown in Ref. \cite{Timsh18}.
$S (f)$ is determined by a correlation character of the analyzed
sequence of the generalized functions. In particular, $S(f)$ can
have dependence of a flicker-noise type: $S (f) \sim 1/f^s$, where
$s \sim 1$. The conclusion about diverse information carried by
the power spectra and the difference moments underlies the FNS
method. Some artificial time series, generalizing the image of the
signal as the sequence of $\delta$-functions, earlier introduced
by Schuster is used in this case.

Any signal, formed exclusively by irregularities-jumps, can be
subject to Fourier transformation to obtain the power spectrum.
Thus, the difference moment $\Phi^{(2)}(t)$ is formed only by
jumps of a dynamic variable at different spatio-temporal levels of
the system hierarchy, and both spikes and jumps contribute to the
power spectrum $S(f)$.

The standard notion about the identity of the information
represented by $\Phi^{(2)}(t)$ and $S(f)$, is valid only for
"smooth" functions. However, real signals $V(t)$ are never smooth.
Therefore, the FNS method focuses on giving the essence of
information on sequences of irregularities latent in real signals
and eliminates information discrepancy, making it possible to
extract the information by considering various features.

Eq. (9) can be used to find the phenomenological parameters $[H_1,
T_1, \sigma]$. Contribution $S_J (f)$ to the power spectrum $S
(f)$, determined by the influence of irregularities, such as
jumps, is expressed by the formula: \be S_J (f)\approx S_J (0)
\frac{1}{1+(2\pi T_1 f)^{2H_1+1}}; \ee \be \nonumber S_J
(0)=4\sigma^2 T_1 H_1 \left\{1-\frac{1}{2H_1 \Gamma^2
(H_1)}\int_0^\infty \Gamma^2(H_1,\xi)d\xi\right\}.\ee Due to
spikes the contribution $S_B(f)$  to power spectrum $S(f)$ can be
generally presented as the expression: \be S_B(f)\approx\frac{S_B
(0)}{1+(2\pi f T_{00})^{n_0}},\ee where parameter $n_0$
characterizes the velocity of the "losses of memory" (correlation
relations) in a sequence of spikes during the time intervals
shorter than the correlation time $T_{00}$. Parameters $ [ n_0,
T_{00}, S_B (0) ] $ characterize self-similarity in the
correlation relations of the peaks. For the resulting power
spectrum an approximation is used: \be
S(f)\approx\frac{S(0)}{1+(2\pi f T_0)^n},\ee where $S (0)$, $T_0$
and $n$ are phenomenological parameters. The parameters which are
determined in such a way differ from the parameters which are used
in Eq.(10): $ S_J (0) \neq S_B (0), T_1 \neq T_0$ and $2H_1 + 1
\neq n_0$. So, the parameters determined from the power spectra
and the structural functions of the second order give different
information. The comparison of the parameters obtained by the
analysis of the experimental series with numerical values of the
parameters for model cases (Fick diffusion, Levy diffusion,
Kolmogorov turbulence, turbulent diffusion, see \cite{Timsh05})
allows to estimate qualitatively the character of the studied
evolution (see \cite{Timsh05} for more information).

The use of irregularities such as spikes, jumps, discontinuities
of derivatives as the information basis of the FNS method allows
to classify and extract phenomenological information contained in
chaotic series. However, evolution of real biological systems has
a more complex and non-stationary nature. In particular,  the
resonance frequencies in the dependences described above, can be
specific for the studied system. The resonance frequencies appear
as peaks in the power spectra $S(f)$ and the oscillatory character
of function $\Phi^{(2)}(t)$. Thus, the values of resonance
frequencies can change during the non-stationary evolution.
Therefore for each state of the studied dynamics, the dependences
$S(f)$ and $\Phi^{(2)}(t)$ are to be considered as "patterns" or
"cliche". These dependences allow to estimate individual
informational characteristics of the state of the system: times of
loss of correlation relations, sets of specific frequencies,
factors of non-stationarity. This will be demonstrated below in
the analysis of Parkinsonian tremor dynamics.

\subsection{Relaxation smoothing of signal, splitting into
"low frequency" and "high frequency" components}

In the analysis of the experimental series we frequently face the
problem of smoothing of the initial signals. Usually the smoothing
polynoms and wavelets are used to filtrate a signal and to extract
a low-frequency component. Here we will describe briefly the
method \cite{Timsh05} of splitting the signal into "low-frequency"
$x_R(t)$ and "high-frequency" $x_F(t)$ components. The method is
based on an iterative procedure of a numeric solution of a heat
conductivity equation: \be \frac{\partial x}{\partial \tau}=\chi
\frac{\partial^2 x}{\partial t^2}.\ee It uses an elementary
explicit finite difference scheme: \be
\frac{x_k^{j+1}-x_k^j}{\Delta
\tau}=\chi\frac{x_{k+1}^j+x_{k-1}^j-2x_k^j}{(\Delta t)^2},\ee that
gives: \bs \be x_k^{j+1}=x_k^j+\frac{\chi\Delta\tau}{(\Delta
t)^2}\left( x_{k+1}^j+x_{k-1}^j-2x_k^j\right).\ee Using
designation $D=\chi\Delta\tau / (\Delta t)^2$, where $D$ is a
diffusion coefficient, we shall rewrite the last equation as: \be
x_k^{j+1}=D x_{k+1}^j+D x_{k-1}^j+(1-2D)x_k^j.\ee Iterations of
this formula give a stable solution to $D < 1/2 ~$ \cite{Timsh05}.
For the smoothing procedure it is necessary to set limiting
conditions. Let smoothing be carried out for a series of $M$
samples (points). At each step of iterations extreme values for $k
= 1$ and $k = M$ are calculated as: \be
x_1^{j+1}=(1-2D)x_1^j+2Dx_2^j, ~~
x_M^{j+1}=(1-2D)x_m^j+2Dx_{M-1}^j.\ee \es When calculating new
values of signal $x_k^{j+1}$ for $j+1$ "relaxation" step by the
values $x_k^j$ (for $j=0$ the initial signal $x(t)$ is set) we can
obtain "low-frequency" component $x_R$. "High-frequency" component
$x_F$ is obtained by subtracting it from the initial signal.
Actually, the smoothing procedure corresponds to the consecutive
reduction of local gradients of values of variables with mutual
rapprochement of points. The splitting of the initial signal
$x(t)$ into two components $x_R(t)$ and $x_F(t)$ enables us to
calculate the dependences $S(f)$ and $\Phi^{(2)}(t)$ for three
functions $x_J(t)~~ (J = R, F$ or $G=R+F)$. Index $G$ is used for
the initial signal $x(t)$. In particular, the low-frequency
("flicker-noise") component, which is present in any chaotic
signal, is  effectively removed when calculating dependence
$S_F(f)$. Therefore, specific frequencies of the studied system
come to light. This will be shown below in the analysis of the
fluctuations of Parkinsonian tremor velocity. The analysis of
dependences $S(f)$ and $\Phi^{(2)}(t)$, which are calculated
separately for each of components $R$ and $F$, presents special
interest for the study of medical parameters.

\section{Analysis of experimental data. Velocity of pathological
tremor in patients with Parkinson's disease}

As the experimental data we use the time records of pathological
tremor velocity in an index finger of patients with Parkinson's
disease \cite{Beuter4,Beuter5}.

Sixteen subjects with Parkinson's disease participated in the
study. All subjects were receiving chronic stimulation either uni-
or bilaterally to relieve the symptoms of Parkinson's disease
including tremor, dyskinesia or rigidity. The participants
received DBS (deep brain stimulation) of the internal globus
pallidus (GPi) or subthalamic nucleus (STN), or ventrointermediate
nucleus of the thalamus (Vim). They were all under 70 years of age
and the group included 11 males and 5 females. All participants
were clinically stable at the moment of the tests; they did not
show cognitive impairment and did not suffer from a major
depressive disorder. All subjects were under minimum dopaminergic
therapy (ranging from 300 to 1200 mg per day of L-Dopa) at the
time of the study and took other Parkinson's disease related
medications.

The selected subjects were asked to refrain from taking their
medication for at least 12 h before the beginning of the tests and
were allowed to have not more than one coffee for breakfast on the
two testing days. Rest tremor was recorded on the most affected
side with a velocity-transducing laser \cite{Beuter6,Norman}. This
laser is a safe (Class II) helium-neon laser. The laser beam is
split, with one part directed at the finger and the other, called
the reference, directed at a rotating disk inside the laser. Back
scattered light from the rotating disk is used to determine the
sign of the velocity signal. Finger tremor was detected and
converted into a calibrated voltage output proportional to finger
velocity. Velocity is more sensitive to low frequency components
inherent in pathological and physiological tremor \cite{Norman}.
Than acceleration is the system did not require any special
calibration procedure. The laser was placed at about 30 cm from
the index finger tip and the laser beam was directed perpendicular
to a piece of a reflective tape placed on the finger tip.

Tremor was recorded with a velocity laser under two conditions of
DBS (on-off), under two conditions of medication (L-Dopa on-off)
and under four conditions of 15, 30, 45, 60 minute periods after
stopping DBS. The conditions, counterbalanced among the subjects
included the following.

1. The "OFF-OFF" condition (no medication and no stimulation).

2. The "ON-ON" condition (on medication and on stimulation).

3. The DBS condition - the "ON-OFF" condition (stimulation only).

4. The L-Dopa condition - the "OFF-ON" condition (no stimulation).

5-8. The "15 OFF", "30 OFF", "45 OFF", "60 OFF" conditions - the
patient's states 15, 30, 45, 60, minutes after the DBS is switched
off, no medication.

\section{Discussion of results. Comparison of statistical
theory of discrete non-Markov random processes and Flicker-noise
spectroscopy}

In this section we present the results of the analysis of the
experimental data \cite{Beuter4,Beuter5,Beuter6}, which are
obtained on the basis of the statistical theory of discrete
non-Markov stochastic processes and the flicker-noise
spectroscopy. As an demonstrative example, we describe the results
for one of the patients (the sixth patient, a woman, 61 years old,
deep brain stimulated bilaterally target structures: subthalamic
nucleus, total daily medication 300 mg). The analysis the of
experimental data allows to reveal some dynamic properties of
Parkinson's disease tremor in each individual case and a group of
patients. The obtained results characterize the group of sixteen
patients in a general way.

\subsection{Non-Markovity, randomness, dynamic alternation and
pathological tremor caused by Parkinson's disease}

\begin{figure}

\includegraphics[width=9cm,height=12cm,angle=270]{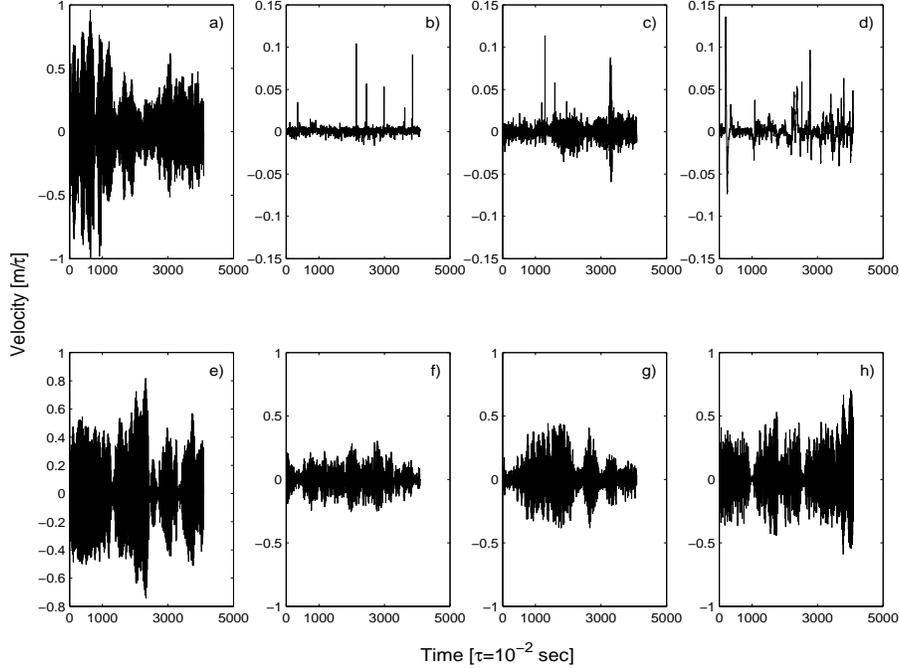}
\caption{Pathological tremor velocity in the left index finger of
the sixth patient with Parkinson's disease. The registration of
Parkinsonian tremor velocity is carried out in the following
conditions: (a) "OFF-OFF" condition (no treatment), (b) "ON-ON"
condition (using DBS and medicaments), (c) "ON-OFF" condition (DBS
only), (d) "OFF-ON" condition (medicaments (L-Dopa) only), (e) -
(h) the "15 OFF", "30 OFF", "45 OFF", "60 OFF" conditions - the
patient's states 15 (30, 45, 60) minutes after the DBS is switched
off, no treatment. Let's note the scale of the pathological tremor
amplitude (see the vertical scale). Such representation of the
time series allows us to note the increase or the decrease of
pathological tremor.}
\end{figure}

In Fig. 1 the initial time series of Parkinsonian tremor velocity
of the patient's (sixth patient) index finger tremor is shown. By
this record it is possible to reveal great differences (see the
vertical scale) in tremor velocity of the patient's state, when
the treatment is not used (Fig. 1a) and under medical treatment
(Figs. 1b-d). The amplitudes of the tremor velocity for the
"OFF-OFF" patient's state (Fig. 1a) and for the "ON-ON" state
(Fig. 1b) differ by 94 times on average. The amplitude of tremor
velocity when the DBS is switched off ("15 OFF", "30 OFF", "45
OFF", "60 OFF" conditions) (Figs. 1e-h) has a residual periodic
character. The analysis of the initial time records does not allow
to determine a method with the best medical effect. In some cases
(Figs. 1e, h) a medical method produces a negative effect that
results in the amplitude increase of tremor velocity and
deterioration of the state of the patient. It is difficult to draw
a conclusion about the efficacy of this or that treatment and to
explain the after-effect of the DBS based on the initial time
signals.

\begin{figure}
\includegraphics[width=9cm,height=12cm,angle=270]{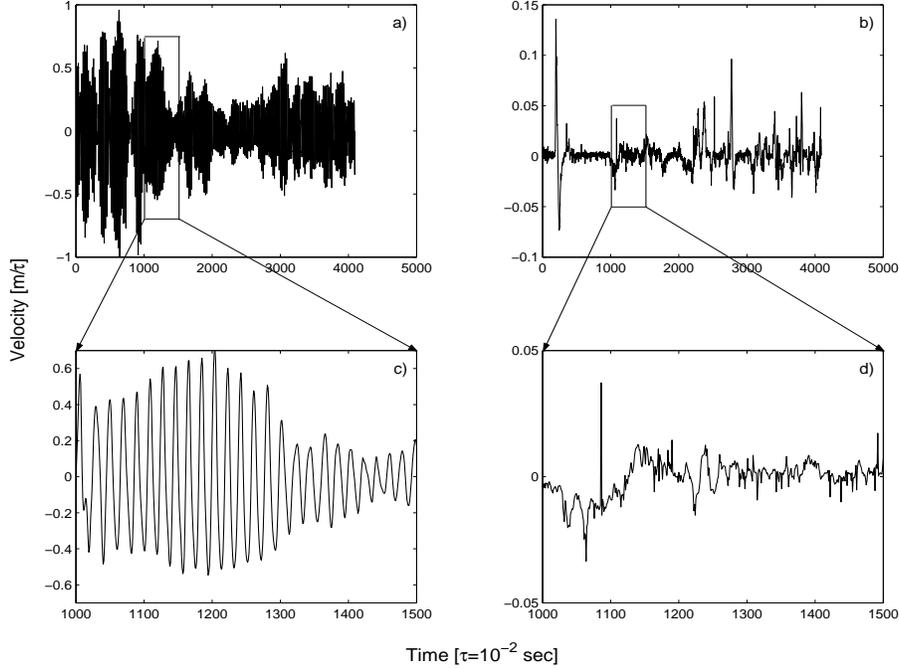}
\caption{The examples of the time series of some areas of
pathological tremor velocity in the sixth patient. (a) The
patient's state with no treatment. (b) The patient with treatment
(medicaments only). The length of local sites of the time series
constitutes 500 points. The local representation clearly reflects
the structure of the initial time signal. (c) The regularity and
the periodic nature of pathological tremor in the patient's state
with no treatment, (d) the randomness of the tremor time signal of
the patient under medicamentous treatment.}
\end{figure}

A different representation of the signal allows to reveal the
effects of periodicity and randomness in the initial time series.
In Fig. 2 the initial time series of tremor velocity in two states
of the patient (see Figs. 1a, d) are shown. We have chosen some
local areas (500 points) of these time records. For the "OFF-OFF"
patient's state (Fig. 2c) the initial time signal has a periodic
structure. Similar periodicity of the initial time signal is
connected to pathological tremor of the patient's limbs (with
frequency $\nu=5.2 Hz$). Retention of the structure is also
observed when the DBS is switched off (see Figs. 1e-h). Medication
produces a different effect on the patient ("OFF-ON" state, see
Fig. 2d). The periodicity of the initial signal is replaced by
randomness. The similar picture is characteristic of all methods
of treatment (see Figs. 1b-d). This conclusion confirms the
general idea about the transition of stochastic dynamic modes in
case of the patient's normal physiological state to the periodic
modes in a pathological state. This reasoning was confirmed by the
analysis of the initial time series of all sixteen patients.

\begin{figure}
\includegraphics[width=9cm,height=12cm,angle=270]{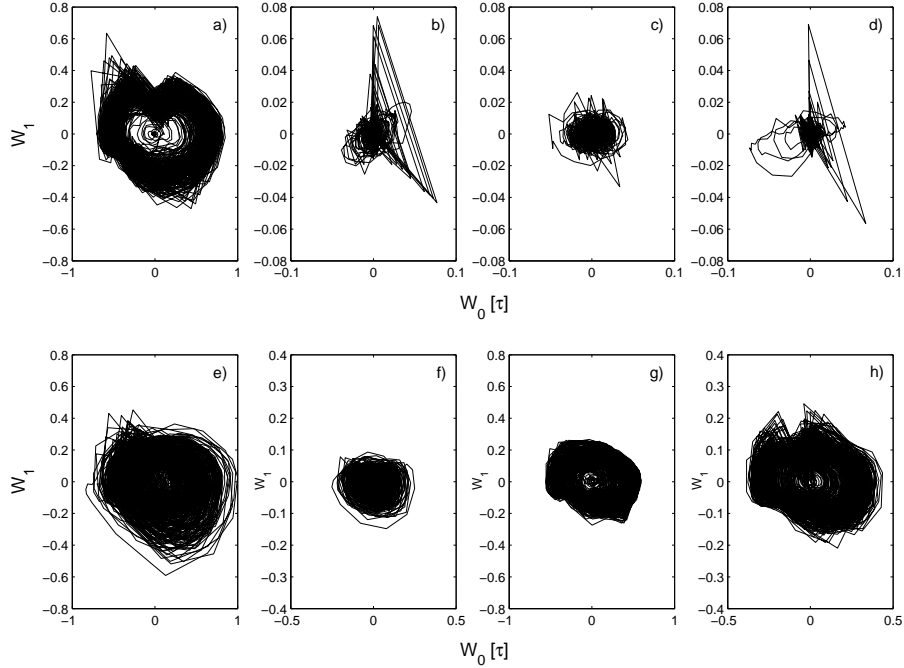}
\caption{The phase portraits of the two first orthogonal dynamic
variables $\textbf{W}_0,\textbf{W}_1$   for pathological tremor in
the patient with Parkinson's disease. The phase portraits are
submitted according to the initial time series. The form of the
phase portraits reflects the initial time signal dynamics. The
periodic structure of the initial signal is determined by
helicoidal phase trajectories (a, e-h); the randomness of the
initial signal is reflected in the curvature of the phase
trajectories (b-d). The most significant fluctuations of the time
signal cause deformation of the phase portraits under any medical
influence. The similar forms of phase clouds are characteristic of
all the groups of patients.}
\end{figure}

The plane projections of phase trajectories of the two first
dynamic orthogonal variables $\textbf{W}_0,~\textbf{W}_1$ (see
Sec. II) in various states of the patient are shown in Fig. 3. All
figures are submitted according to the initial time series. The
structure of the phase trajectories is determined by the presence
of fluctuations in the initial time series. The most significant
fluctuations lead to deformation of phase clouds. The phase
portraits consist of empty cores and helicoidal trajectories with
a high concentration of phase points in the "OFF-OFF" patient's
state (Fig. 3a) and in the cases when the DBS is switched off ("15
OFF", "30 OFF", "45 OFF", "60 OFF", see Figs. 3e-h). The presence
of empty cores is explained by the insignificant quantity of small
fluctuations near a zero value. Such structure of the phase
portrait is determined by periodicity in the initial time signal
(Fig. 2c). Here one can note a completely different picture of the
effect any medical treatment has on the patient's organism. One
can see rare significant fluctuations instead of small
fluctuations of tremor velocity. These peak deviations result in
acute-angled deformations of the central cores with a high
concentration of phase points. The form of phase portraits
corresponds to the dynamic alternation of periodic and stochastic
components of the initial time signal. It is possible to draw a
conclusion about the most effective method of treatment of each
patient judging by the form of phase clouds. For the sixth patient
two methods of treatment are almost equivalent: the complex
treatment of a patient by two medical methods (Fig. 3b) and by
using medicaments only (Fig. 3d).

\begin{figure}
\includegraphics[width=9cm,height=12cm,angle=270]{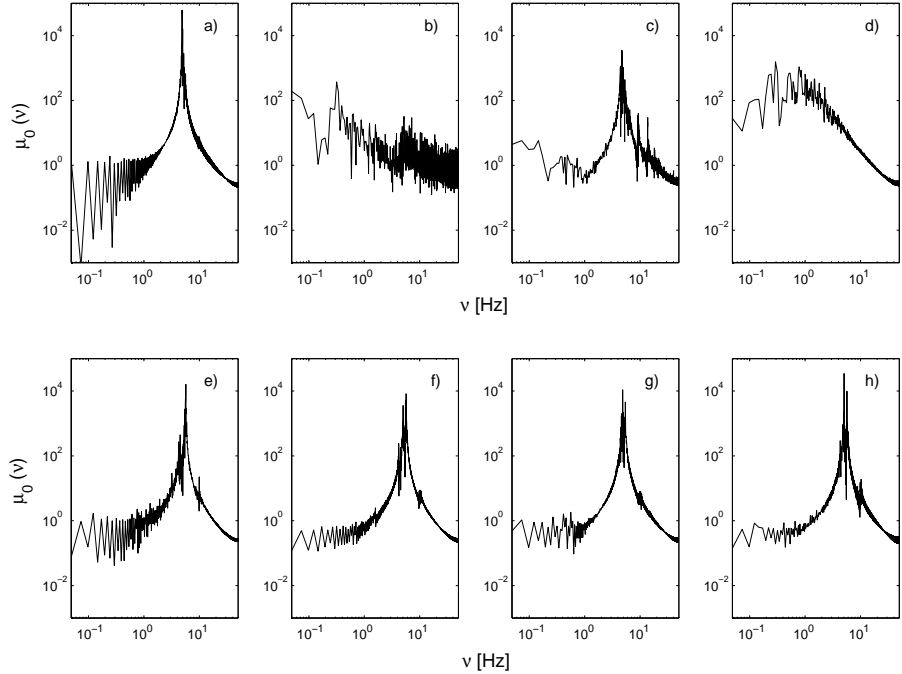}
\caption{The initial time correlation function $\mu_0(\nu)$  power
spectrum for pathological tremor in the patient with Parkinson's
disease. The power spectra are submitted according to the
arrangement of the initial time series. The log-log scale is
chosen to receive a more detailed information. In the initial TCF
$\mu_0(\nu)$ power spectrum the power peak is observed on
frequency $\nu=5.2 Hz$. The amplitude of this peak is determined
by the amplitude of the initial time signal. The oscillations (a,
e-h) are clearly expressed in the low frequencies area of the
power spectrum. The low-frequency oscillations vanish, and the
amplitude of the power peak on frequency $\nu=5.2 Hz$ decreases
with the application of any method of treatment (b-d). The
amplitude of this peak changes a little bit when the DBS is used.
It reflects low efficiency of this method of treatment (of the
sixth patient). The form of power spectrum $\mu_0(\nu)$  is
determined by the nature of the initial time signals.}
\end{figure}

The initial TCF power spectrum $\mu_0 (\nu)$ (see Sec. II) in
various physiological states of the patient is shown in Fig. 4.
All dependences are presented on a log-log scale. At
characteristic frequency $\nu=5.2 Hz$ the peak is observed for
each power spectrum. The peak amplitude is determined by the
amplitude of tremor velocity in the initial time signal. In case
of any medical treatment the peak amplitude considerably
decreases. It happens because the amplitude of pathological tremor
decreases. Any power spectrum reflects dynamic alternation effects
in the initial time series. Oscillations are seen in the power
spectra (Figs. 4a, e-h). Their periodic nature is clearly
expressed by a low frequency range. The periodic structure of the
power spectra is broken, oscillations disappear, the amplitude of
the peak on frequency $\nu=5.2 Hz$ decreases under treatment. Such
form of the power spectra is connected to the amplification of
randomness effects in the initial time series. The periodic nature
of the patient's Parkinsonian tremor velocity is replaced by
stochastic fluctuations with a low amplitude of tremor velocity.
On the whole, the comparative analysis of the power spectra shows
that medicamentous treatment has the most significant effect on
the patient's organism (thus the combined use of medicaments and
DBS is almost equivalent). The use of DBS stimulation does not
decrease abnormal oscillations significantly.

The effects of randomness and regularity are most visible in the
frequency dependences of the first point of the non-Markov
parameter $\varepsilon_1 (\nu)$. The physical idea of this
parameter consists in revelation of Markov and non-Markov features
in the time series. This parameter classifies all stochastic
processes into random Markov processes $\varepsilon \gg 1$,
quasi-Markov (intermediate) processes $\varepsilon >1$ and
non-Markov processes $\varepsilon \sim 1$. Thus, the first point
of the non-Markovity parameter at zero frequency $\varepsilon_1
(0)$ contains the most significant information. This point
accumulates the information about all dynamic peculiarities of the
time series. The increase of this value testifies to the
amplification of Markov random effects in the time signals and the
appearance of the effects of short-range or instant memory. Thus,
the comparative analysis of various physiological states of the
patient shows that the higher value of $\varepsilon_1 (0) \sim
10^1$ corresponds to the smaller pathological tremor velocity. For
the smallest pathological tremor velocities (in one of the
patients of the studied group) the values of this parameter are
maximal and achieve the value of $\varepsilon_1 (0) \sim 10^2$. On
the contrary, the decrease of this parameter reflects the
amplification of non-Markov effects in the initial time series.
Thus, the decrease of parameter $\varepsilon_1 (0)$ corresponds to
the increase in Parkinsonian tremor velocity in the initial time
signal. The greatest pathological tremor velocities (for one of
the patients of the studied group) are characterized by minimal
values $\varepsilon_1 (0) \sim 10^0$. Thus, the value of
$\varepsilon_1 (0)$ reflects the behavior of the patient's tremor
amplitude.

It should be noted that we use the value of the first point of the
non-Markovity parameter at zero frequency $\varepsilon_1 (0)$ as
the quantitative measure which reflects Markov and non-Markov
effects in the initial time series. The values of the
non-Markovity parameter at zero frequency determine long-range
correlations (in a limiting case for $t\rightarrow\infty$). The
greater values of parameter $\varepsilon_1 (0)$ determine stronger
correlations. This thesis confirms a general concept that
long-range correlations should take place in a stable healthy live
subject. We remind that in this work we analyze the time series of
pathological tremor of a patient and we observe only minimal
alterations of parameter $\varepsilon_1 (0)$  (from a unit up to
several tens.)

\begin{figure}
\includegraphics[width=9cm,height=12cm,angle=270]{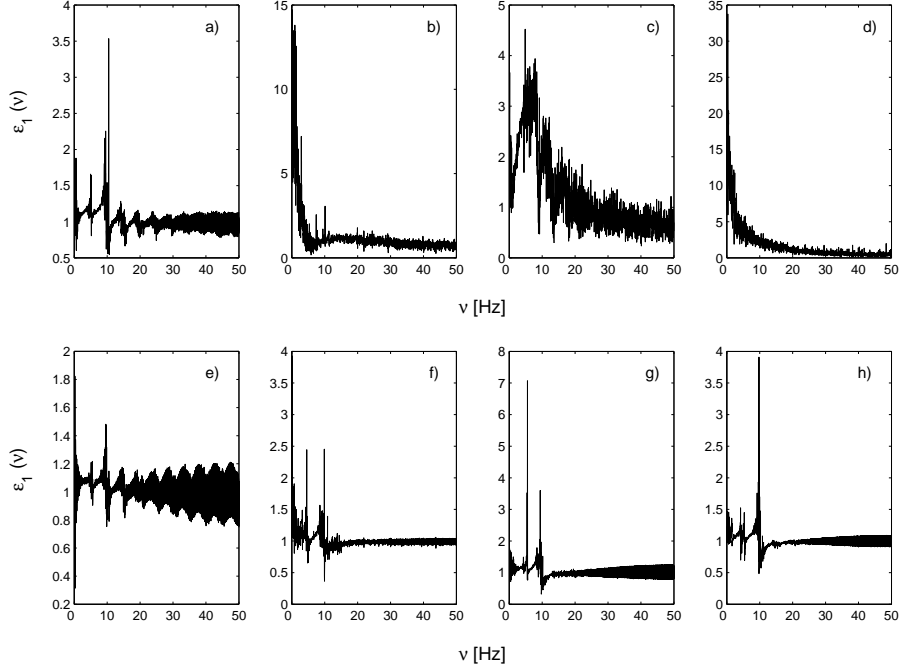}
\caption{The frequency dependence of the first point of the
non-Markovity parameter $\varepsilon_1(\nu)$ for pathological
tremor velocity in the patient. For example, the sixth patient
with Parkinson's disease is chosen. The figures are submitted
according to the arrangement of the initial time series. The
characteristic low-frequency oscillations are observed in
frequency dependences (a, e-h), which get suppressed under medical
influence (b-d). The non-Markovity parameter reflects the Markov
and non-Markov components of the initial time signal. The value of
the parameter on zero frequency $\varepsilon_1(0)$ reflects the
total dynamics of the initial time signal. The maximal values of
parameter $\varepsilon_1(0)$ correspond to small amplitudes of
pathological tremor velocity. The minimal values of this parameter
are characteristic of significant pathological tremor velocities.
The comparative analysis of frequency dependences
$\varepsilon_1(\nu)$ allows us to estimate the efficiency of each
method of treatment.}
\end{figure}

The periodic structure of pathological tremor velocity is
reflected also in frequency dependences of the first point of the
non-Markovity parameter (Figs. 5a, e-h). Oscillations emerge with
a characteristic frequency $\nu \sim 5.2 Hz$. The oscillations are
most appreciable in the low frequency range. This structure is
completely suppressed by medical influence on the patient's
organism. The fast change of various dynamic modes and the
amplification of randomness effects result in the infringement of
the periodic picture. It is connected with the decrease of
pathological tremor, the "ON-ON", "OFF-ON" states (Figs. 5b, d).

The analysis of the values of parameter $\varepsilon_1 (0)$  in
different  states of the patient specifies the most effective
method of treatment (pathological tremor suppression). Values
$\varepsilon_1 (0)_M=14.87$ in the "OFF-ON" patient's state (using
medicaments) and $\varepsilon_1 (0)_{ON}=13.53$ in the "ON-ON"
patient's state (using medicaments and the DBS) show, that these
methods of treatment have almost the same positive effects. For
comparison: we have $\varepsilon_1 (0)_{DBS}=3.56$ in the "ON-OFF"
state of the studied patient (using the DBS), and $\varepsilon_1
(0)_{OFF}=1.47$ in the "OFF-OFF" state of the patient (no
treatment).

\begin{figure}
\includegraphics[width=9cm,height=12cm,angle=270]{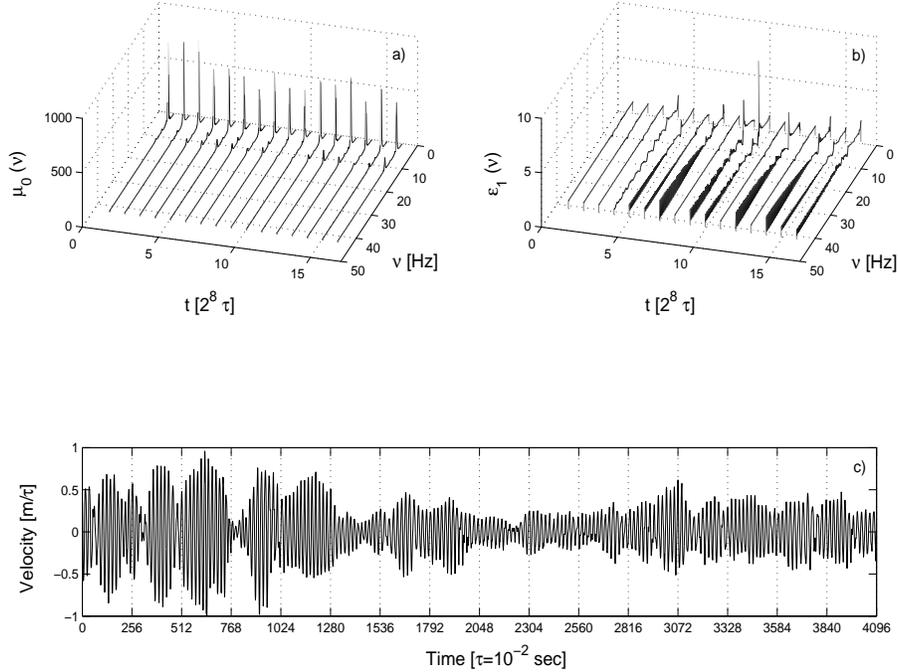}
\caption{The window-time behavior of the power spectrum
$\mu_0(\nu)$ (a) and the frequency dependence of parameter
$\varepsilon_1(\nu)$ (b) for pathological tremor in the "OFF-OFF"
condition (c). The local representation of the TCF power spectrum
and the frequency dependence of the -non-Markovity parameter
allows us to reflect the dynamic features of some areas of the
time series. The localization procedure consists in choosing the
optimum length of the local sample, in dividing the time series
and in designing local characteristics for each sample. The
increase of pathological tremor velocity results in the increase
of the amplitude of the power peaks at a characteristic frequency
and in the decrease of the non-Markovity parameter to a unit.}
\end{figure}

The window-time behavior of the initial TCF power spectrum
$\mu_0(\nu)$ (Fig. 6a) and the first point of the non-Markov
parameter $\varepsilon_1 (\nu)$ (Fig. 6b) is shown in Fig. 6. For
example, we selected the initial time series in the "OFF-OFF"
patient's state. The design of this dependence was realized by
means of the localization procedure. The peculiarity of this
procedure consist in reflecting the dynamical properties of the
local area of the initial time signal \cite{Yulm6}. First, the
optimal length of the sample (window) must be determined. The
preliminary analysis of various window lengths gave the optimal
length of $256=2^8$ points. The correlation function power
spectrum $\mu_0(\nu)$  and the non-Markovity parameter
$\varepsilon_1 (\nu)$ are calculated for each of the windows. This
procedure shows the dynamic features of local areas of the initial
time signal.

In Fig. 6a power peaks are observed at frequency $\nu=5.2 Hz$ in
the window-time behavior of the initial TCF power spectrum. The
amplitude of these peaks reflects the increase or decrease in
Parkinsonian tremor. In particular, the most significant peaks are
visible for windows 1-3. Thus, the highest pathological tremor of
the initial time record corresponds to these areas. The similar
behavior of the power spectrum confirms the periodic nature of the
initial time signal. The behavior of the non-Markovity parameter
$\varepsilon_1(\nu)$ in Fig. 6b is as follows. The value of
parameter $\varepsilon_1(0)$ reaches a unit (see, windows 1-3, 8,
11, 13) at the moments of the increase in pathological tremor
velocity. Thus, the value of the non-Markovity parameter starts to
decrease 2.5-3 sec before the increase in tremor velocity. When
pathological tremor velocity decreases, the value of parameter
$\varepsilon_1(0)$ increases to values 2-3 (in the "OFF-OFF"
patient's state, no treatment).

In Table 1 we present the root-mean-square amplitude $\langle A
\rangle=\{\sum_{j=0}^{N-1} x_j^2/N\}^{1/2}$, and the dispersion
$\sigma^2=(1/N){\sum_{j=0}^{N-1} (x_j-\langle X \rangle)^2}$ of
kinetic parameter $\lambda_1$ for the sixth patient. The physical
meaning of parameter $\lambda_1$  consists in determining the
relaxation rate of the studied process \cite{Yulm7}. The ratio of
the mean-squared amplitude in the "OFF-OFF" patient's state (no
treatment) and in the "ON-ON" patient's state (using medicaments
and the DBS) makes 250 (!) times. It shows, that after the
application of any type of treatment, relaxation rate increases.
Sharp distinctions of relaxation rate reflect pathological and
normal physiological processes. Thus, this quantitative
characteristic allows to reveal effective or inefficient methods
of treatment (to decrease tremor velocity).

\begin{flushleft}
\footnotesize Table 1
\\ The root-mean-square amplitude $\langle A \rangle$, and dispersion
$\sigma^2$ (absolute values) of the first kinetic parameter
$\lambda_1$ in various physiological states of the sixth patient,
calculated by means of our theory. 1 - DBS, 2 - Medication
(L-Dopa). For example, OFF OFF - DBS off, medication off.
\end{flushleft}

\begin{center}
\footnotesize
\begin{tabular}{p{1cm}p{1.8cm}p{1.8cm}p{1.8cm}p{1.8cm}p{1.8cm}p{1.8cm}p{1.8cm}p{1.8cm}}
\hline  & OFF OFF & ON ON & ON OFF & OFF ON & 15 OFF & 30 OFF & 45
OFF & 60 OFF \\
\hline $\langle A \rangle$ & $17.4\cdot 10^{-5}$ & $43.4\cdot
10^{-3}$ & $21.7\cdot 10^{-3}$ & $52.4\cdot 10^{-3}$
 & $10.7\cdot 10^{-5}$ & $10.6\cdot 10^{-5}$ & $10.4\cdot 10^{-5}$ & $13.8\cdot 10^{-5}$ \\
$\sigma^2$ & $59.3\cdot 10^{-3}$ & $0.392$ & $0.291$ & $0.569$ &
$51.5\cdot 10^{-3}$ & $62.2\cdot 10^{-3}$ & $63.3\cdot 10^{-3}$ & $61.3\cdot 10^{-3}$ \\
\hline
\end{tabular}
\end{center}

\subsection{The use of the flicker-noise spectroscopy for the analysis
of pathological tremor velocity caused by Parkinson's disease}

\begin{figure}
\includegraphics[width=9cm,height=12cm,angle=270]{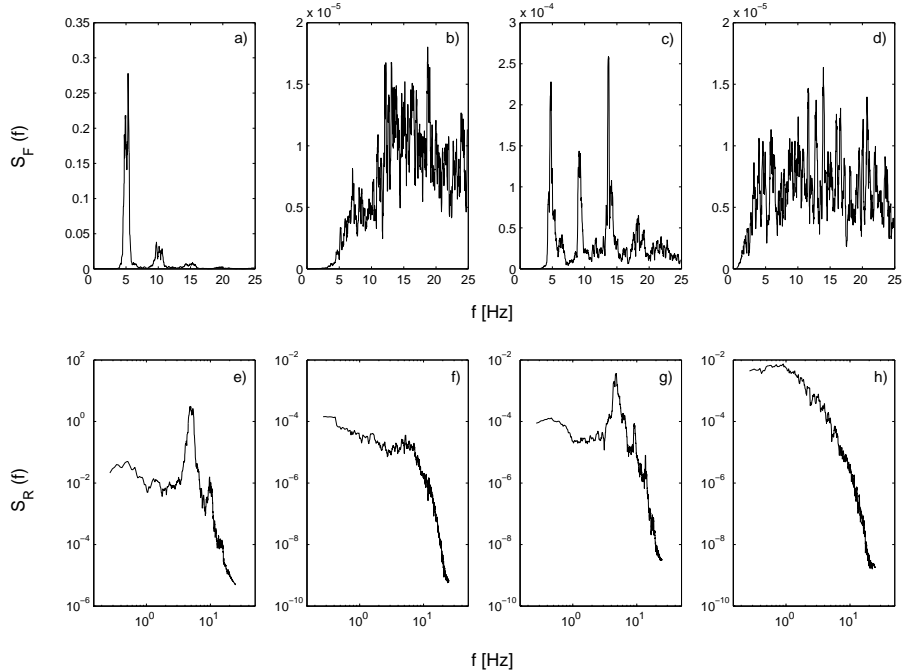}
\caption{The power spectra of high-frequency (a-d) and
low-frequency (e-h) components of the initial time signal in the
"OFF-OFF" condition (a, e) and in various conditions of treatment:
medicaments and the DBS ("ON-ON" condition) (b, f), the DBS only -
("ON-OFF" condition) (c, g), medicaments only ("OFF-ON" condition)
(d, h). The power spectra of high-frequency component $F$ are
submitted on a linear scale (a-d), the power spectra of
low-frequency component $R$ are submitted on a log-log scale
(e-h). The set of eigenfrequencies in the power spectra is
connected to the periodic nature of Parkinsonian tremor.}
\end{figure}

In Fig. 7 the power spectra of high-frequency $S_F(f)$ (Figs.
7a-d) and low-frequency $S_R(f)$ (Figs. 7e-h) components of the
initial time signal (the tremor of the index finger velocity in
the four patient's states, see Figs. 1a-d) are shown. Frequency
dependence of the power spectrum of high-frequency component
$S_F(f)$ is on a linear scale, and the power spectrum of
low-frequency component of time signal $S_R(f)$ is on a log-log
scale. The small set of eigenfrequencies is characteristic of
tremor of the "OFF-OFF" patient's state (Figs. 7a, e). These
frequencies are clearly visible at $f_{01} \sim 5.2 Hz$ and
$f_{02} \sim 10.1 Hz$. According to Figs. 7c, g, the situation
does not change qualitatively, when the DBS ("ON-OFF" patient's
state) is used, as the small set of eigenfrequencies of tremor is
preserved. However, the maximal values of $S_F(f)$ decrease by
three orders and maximal values of $S_R(f)$ decrease by two
orders. The situation changes qualitatively when medicaments are
used ("OFF-ON", Figs. 7d, h), and also in case of the combined use
of the DBS and medicaments ("ON-ON", Figs. 7b, f). Tremor velocity
in these cases becomes more chaotic, since a wide band of
eigenfrequencies appears in the tremor power spectrum. The maximal
values of frequencies in dependence $S_F(f)$ decrease by four
orders, the values of $S_R(f)$ decrease in the low frequencies
range by two orders under the combined use of the DBS and
medicaments ("ON-ON"). Values $S_R (f) $ in the low-frequencies
area (when using medicaments, "OFF-ON") are commensurable with
values $S_R (f) $ in the "OFF-OFF" state.

\begin{figure}
\includegraphics[width=9cm,height=12cm,angle=270]{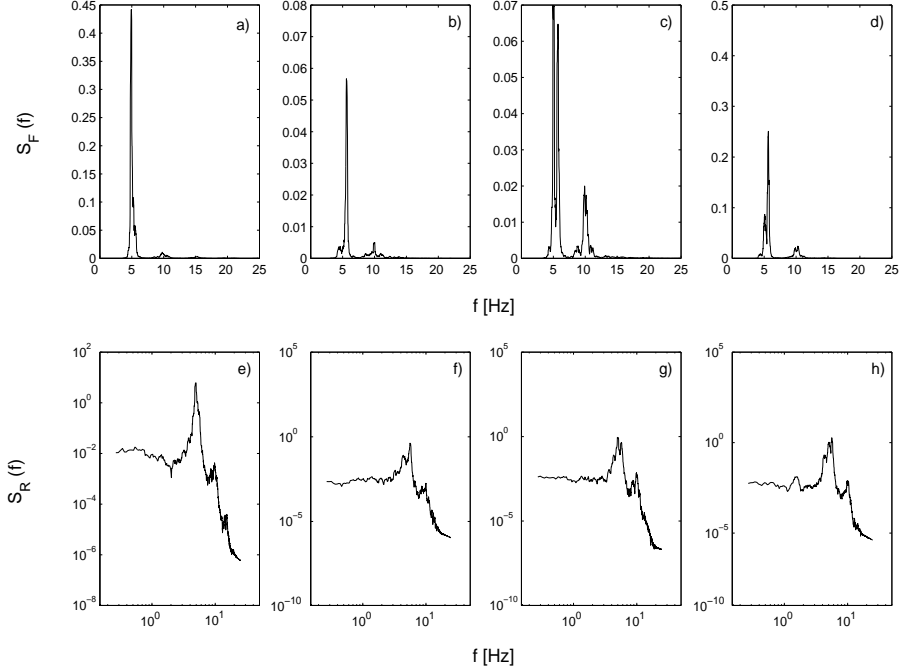}
\caption{The power spectra of high-frequency (a-d) and
low-frequency (e-h) components of Parkinsonian tremor velocity in
the patient's states 15 (a, e), 30 (b, f), 45 (c, g), 60 (d, h)
minutes after the DBS is switched off. The high-frequency
component $F$ is submitted on a linear scale, the low-frequency
$R$ is submitted on a log-log scale. From these figures it is
visible, that by switching off the DBS we get the insignificant
decrease of the resonant component of the initial signal.}
\end{figure}

The specific features characterizing tremor velocity in the
patient with Parkinson's disease are seen in spectra $S_F(f)$ and
$S_R(f)$ in Figs. 7a-h. It should be noted that the splitting of
the initial time signal $x(t)$ into components $x_R(t)$ and $x_F
(t)$ depends on quantity $N$ of iterations of "diffusion"
smoothing and the value of "diffusion coefficient" $D$ ($N=10$ and
$D = 1/4$ for the plots). The analysis shows that the increase of
the number of such iterations up to $N = 50$ for various values of
parameter $D$ does not change the obtained results essentially.
Therefore, we use values $N = 10$ and $D = 1/4$ to calculate FNS
dependences in the following way.

\begin{figure}
\includegraphics[width=9cm,height=12cm,angle=270]{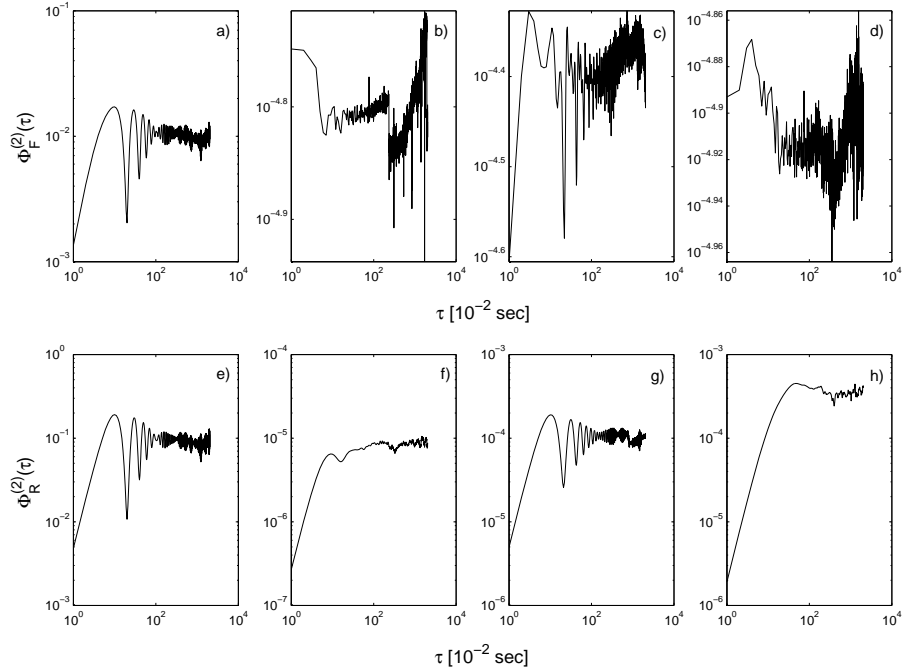}
\caption{The difference moments of the second order for
high-frequency component $F$ and low-frequency component $R$ in
the states of the patient: (a, e) - no treatment ("OFF-OFF"
condition), (b, f)- complex treatment ("ON-ON" condition), (c, g)
- the DBS only ("ON-OFF" condition), (d, h) - medicaments only
("OFF-ON" condition). The figures are submitted on a log-log
scale. The behavior of difference moments $\Phi_F^{(2)}(\tau),
~\Phi_R^{(2)}(\tau)$ becomes more stochastic in any case of
medical influence.}
\end{figure}

Fig. 8 shows the dynamics of tremor velocity of the patient in the
cases when the DBS is switched off. The use of this medical
methods does not change the situation. The small set of
eigenfrequencies of tremor velocity remains the same. However,
there is significant decrease in the maximal values of power
spectra $S_F(f)$ and $S_R(f)$. Frequency dependence $S_F(f)$
should be considered as information valuable patterns when
determining the characteristic frequencies of the system, as
$S_R(f)$ dependence do not changes significantly (see Fig. 8).

The difference moments of low-frequency and high-frequency
components of the initial signal $\Phi_R^{(2)}(\tau)$,
$\Phi_F^{(2)}(\tau)$ can also carry valuable information about
tremor dynamics. It is seen in Figs. 9-10. This dependence
characterizes the duration of correlation interval $T_1$ during
which "a loss of memory" about the local value of the signal takes
place.

\begin{figure}
\includegraphics[width=9cm,height=12cm,angle=270]{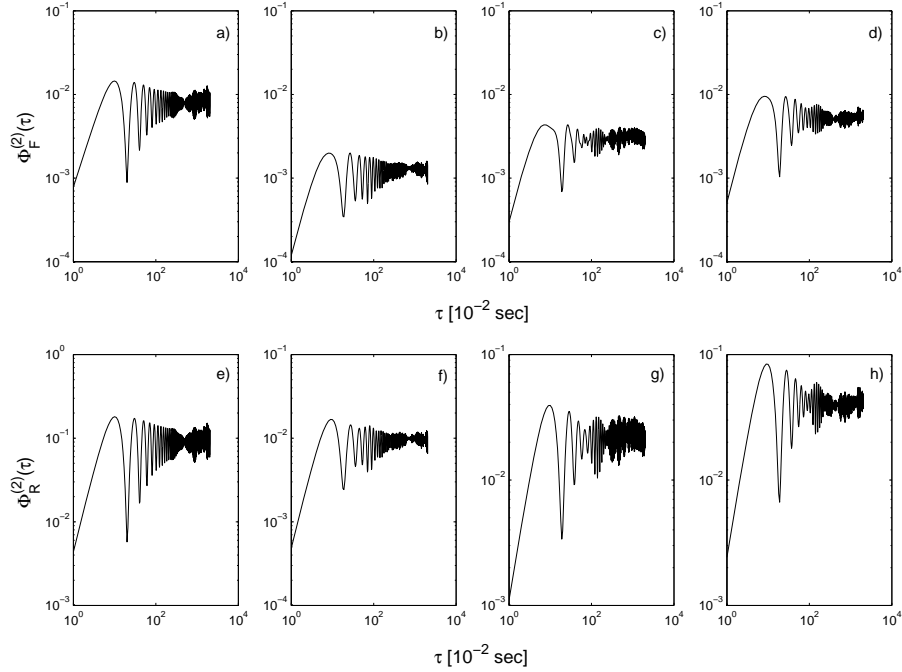}
\caption{The difference moments of the second order for
high-frequency component $F$ and low-frequency component $R$ of
Parkinsonian tremor in the patient: (a, e) 15 minutes after the
DBS is switched off, (b, f) 30 minutes after the DBS is switched
off, (c, g) 45 minutes after the DBS is switched off, (d, h) - 60
minutes after the DBS is switched off. Medicaments are not used.
The figures are submitted on a log-log scale. The aftereffects of
the DBS render insignificant influence on the initial
characteristics of tremor velocity.}
\end{figure}

The combined use of the DBS and medicaments ("ON-ON", Figs. 9b, f)
results in the strongest changes in tremor velocity: the
dispersion of fluctuations decreases by two orders and the
contribution of "resonance" frequencies is considerably
suppressed. It is seen in Fig. 9. The latter conclusion follows
from the decrease of the oscillatory component.

Time dependence in Fig. 10 characterizes the dynamics of
pathological tremor velocity in the patient's states when the DBS
is switched off ("15 OFF", "30 OFF", "45 OFF", "60 OFF"). The
comparative analysis of Figs. 9 and 10 shows that the initial
parameters of the patient's tremor ("OFF-OFF" state) almost do not
vary. Thus, the dispersion of fluctuations somewhat decreases and
the "resonance" component of the initial signal actually does not
change.

\section{Conclusions. The analysis of pathological tremor caused by Parkinson's disease}

In this work we offer a new method of analyzing one of the
symptoms of Parkinson's disease. This method is based on the
combined use of the statistical theory of discrete non-Markov
stochastic processes and the flicker-noise spectroscopy. Each of
the offered theories reflects the specific dynamic peculiarities
of Parkinson's disease tremor. Combined representation of the
obtained results allows to extract the most complete and
trustworthy information about the dynamics of Parkinsonian tremor.

As experimental data we used the time series of the tremor
velocity of an index finger of the patient with Parkinson's
disease for various methods of treatment. The tremor of a human's
limbs certainly is just one of the external symptoms of
Parkinson's disease. However the tremor dynamics of a human's limb
is caused by complex interrelation of separate areas of brain,
central nervous system, locomotor system, chemical and biological
metabolism. Any anomaly of one of these systems is clearly
reflected in tremor dynamics. The analysis of the stochastic
dynamics of a human's index finger tremor is only part of the
complex analysis of the patient's states.

The analysis of stochastic dynamics of the pathological tremor
velocity shows the change of the physical nature of the signal.
The robust periodic structure with specific oscillation frequency
is characteristic of the time series of pathological tremor
velocity in "OFF-OFF" patient's state (no treatment). The similar
structure of the initial time signal is connected to the
pathological tremor of the limbs in the patient with Parkinson's
disease. The structure of the signal changes and becomes more
stochastic when the patient undergoes treatment. The amplification
of stochastic effects in the initial signal is connected to the
increase of degrees of freedom in the studied system. Their
maximal value corresponds to a normal physiological state.

The first technique consists in determining the Markov and
non-Markov components of the initial time signal, manifestation of
long-range and short-range statistical memory effects, the
amplification and degradation of correlations in the initial time
signal. This technique also allows to reveal the effects of
randomness, regularity and periodicity, dynamic alternation
effects of various relaxation processes for overall initial time
series and their local areas. The statistical theory of discrete
stochastic non-Markov processes makes it possible to obtain the
whole spectrum of quantitative and qualitative values and the
characteristics based on the information carried by fluctuations
and correlations in the initial time signals.

The stochastic dynamics of pathological tremor velocity in various
dynamic states of a patient with Parkinson's disease is clearly
reflected in the phase portraits of orthogonal variables, in power
spectra of a correlation function and memory functions of the
lower orders, in frequency dependences of the non-Markov
parameter, in local relaxation and kinetic parameters. The
analysis of all the information about the correlations and
fluctuations of the initial time series allows to estimate the
efficiency of different methods of treatment. For example, we
analyzed the experimental data of one of the patients with
Parkinson's disease.

We used the statistical spectrum of the non-Markovity parameter to
reveal Markov and non-Markov effects in the initial time series.
The increase or decrease of pathological tremor velocity
predetermines the changes of the statistical non-Markovity
parameter. The increase of the stochastic effects and the Markov
components of the initial time signal and the manifestation of
long-range correlations correspond to the decrease of pathological
tremor velocity. At the same time the increase of pathological
tremor velocity is accompanied by the occurrence of the effects of
periodicity and regularity, harmonic oscillations, non-Markov
effects, the manifestation of memory effects and degradation of
correlations in the initial time signals.

The described localization procedure reflects the dynamic
properties of pathological tremor in some local areas of the
initial time series. The window-time behavior of the initial time
correlation function $\mu_0(\nu)$ and the first point of the
non-Markovity parameter $\varepsilon_1(\nu)$ reflects the
predictors of changes in pathological tremor velocity. Any changes
of pathological tremor are instantly reflected in the window-time
behavior of the dynamic characteristics.

Dynamic characteristics $\langle A \rangle$ and $\sigma^2$ of
relaxation parameter $\lambda_1$ reflect relaxation rate in
various dynamic patient's states. The use of any method of
treatment results in acceleration of stabilization process.
Significant distinctions in the dynamic characteristics of
parameter $\lambda_1$ in the state of the patient with no
treatment ("OFF-OFF" condition) and in the complex use of
medicaments and the DBS ("ON-ON" condition) allow to estimate the
relaxation scales of physiological processes for healthy and sick
people.

Finally, the change of the nature of the pathological tremor time
signal results in the change of the whole set of quantitative and
qualitative characteristics and exponents. The general algorithm
of these changes can be determined in the following way. The
structure of the initial time series becomes more stochastic with
reduction of Parkinsonian tremor velocity. The increase of the
number of degrees of freedom determines predominance of Markov
effects, manifestation of long-range correlations and more
significant velocity of relaxation processes. The regularization
of the initial time signal is observed with the increase of
pathological tremor velocity (when this or that treatment is
stopped), i.e. its structure becomes more robust. This fact is
reflected in the behavior of the physical characteristics. The
structure of phase clouds changes, the height of the peak on the
characteristic frequency in the initial TCF power spectra
increases, characteristic oscillations in the frequency spectra of
the non-Markovity parameter appear, the value of the first point
of the non-Markovity parameter on zero frequency decreases to
$~1$.

The use of the flicker-noise spectroscopy confirms the conclusion
about the resonant character of the amplitude of Parkinsonian
tremor velocity. It is necessary to note that the influence of the
resonant tremor component decreases, and the stochastic tremor
component increases under treatment. At the same time the analysis
of spectral dependences and different moments of the second order,
as well as the splitting of the signal into "high-frequency" and
"low-frequency" components have allowed to make the timely
estimate of the patient's state (Figs. 8, 10) and the efficiency
of different methods of treatment (Figs. 7, 9). The estimates,
which we have received, can be used to choose the "treatment
strategy" for the patient. It is necessary to note, that the
high-frequency component of pathological tremor velocity $S_F (f)$
is most sensitive to the changes arising in the initial time
signal. On the whole, the method of the flicker-noise spectroscopy
allows to reveal the set of the eigenfrequencies and the resonant
or the stochastic components of the initial time signal in various
physiological states of the patient. This method elicits
additional information about the effects of dynamic alternation
and the nonstationary effects in the initial time series.

The methods of the time series analysis which are submitted in
this paper, give a simple quantitative method or a graphic scheme
for the analysis of various physiological states of a patient with
Parkinson's disease. Each of these methods is independent and
reflects unique local information about the initial time signal
structure.

The obtained results allow us to use the offered methods for the
analysis of other complex systems of biological nature. In
particular, the offered methods have already made it possible to
receive significant results in the statistical analysis of the
experimental data of epidemiology, cardiology, neurophysiology and
of human locomotion.

\section{Acknowledgements}

This work was supported in part by the RFBR (Grants no.
05-02-16639-a, 04-02-16850, 05-02-17079), RHSF (Grant no.
03-06-00218a) and Grant of Federal Agency of Education of Ministry
of Education and Science of RF. This work has been supported in
part (P.H.) by the German Research Foundation, SFB-486, project
A10. The authors acknowledge Professor, Dr. Anne Beuter and Dr.
J.M. Hausdorff for stimulating criticism and valuable discussions,
Dr. L.O. Svirina for technical assistance.

\end{document}